\documentclass[prd,twocolumn,unsortedaddress,superscriptaddress]{revtex4}
\usepackage{graphicx}
\usepackage{dcolumn}
\usepackage{amsmath}

\newcommand{\Kpen}{\ensuremath{K_{L} \rightarrow \pi^{\pm} e^{\mp} \nu_{e}}}
  \newcommand{\kpen}{\ensuremath{K_{e3}}}
\newcommand{\Kpeng}
  {\ensuremath{K_{L} \rightarrow \pi^{\pm} e^{\mp} \nu_{e} \gamma}}
  \newcommand{\kpeng}{\ensuremath{K_{e3\gamma}}}
\newcommand{\Kpmz} {\ensuremath{K_{L} \rightarrow \pi^+ \pi^- \pi^0}}
  \newcommand{\kpmz}{\ensuremath{K_{\pi3}}}

  \newcommand{\kpmn}{\ensuremath{K_{\mu 3}}}
\newcommand{\Kpm} {\ensuremath{K_{L} \rightarrow \pi^+ \pi^-}}

  \newcommand{\kppen}{\ensuremath{K_{e4}}}
\newcommand{\Egs} {\ensuremath{E_\gamma^*}}
\newcommand{\thegs} {\ensuremath{\theta_{e\gamma}^*}}
\newcommand{\MeV} {\ensuremath{\rm MeV}}
\newcommand{\Acc}{\ensuremath{\rm Acc}}
\newcommand{\BR}{\ensuremath{\rm BR}}
\newcommand{\Num}{\ensuremath{\rm N}}

\begin{document}

\title{A New Measurement of the Radiative \kpen\ 
Branching Ratio and Photon Spectrum}

\affiliation{University of Arizona, Tucson, Arizona 85721}
\affiliation{University of California at Los Angeles, Los Angeles,
  California 90095} 
\affiliation{University of California at San Diego, La Jolla,
  California 92093} 
\affiliation{The Enrico Fermi Institute, The University of Chicago, 
  Chicago, Illinois 60637}
\affiliation{University of Colorado, Boulder, Colorado 80309}
\affiliation{Elmhurst College, Elmhurst, Illinois 60126}
\affiliation{Fermi National Accelerator Laboratory, Batavia, Illinois 60510}
\affiliation{Osaka University, Toyonaka, Osaka 560-0043 Japan}  
\affiliation{Rice University, Houston, Texas 77005}
\affiliation{Rutgers University, Piscataway, New Jersey 08854}
\affiliation{The Department of Physics and Institute of Nuclear and 
Particle Physics, University of Virginia, 
Charlottesville, Virginia 22901}
\affiliation{University of Wisconsin, Madison, Wisconsin 53706}

\author{A.~Alavi-Harati}
\affiliation{University of Wisconsin, Madison, Wisconsin 53706}
\author{T.~Alexopoulos}
\affiliation{University of Wisconsin, Madison, Wisconsin 53706}
\author{M.~Arenton}
\affiliation{The Department of Physics and Institute of Nuclear and 
Particle Physics, University of Virginia, 
Charlottesville, Virginia 22901}
\author{K.~Arisaka}
\affiliation{University of California at Los Angeles, Los Angeles,
  California 90095} 
\author{S.~Averitte}
\affiliation{Rutgers University, Piscataway, New Jersey 08854}
\author{R.F.~Barbosa}
\altaffiliation[Permanent address ]{University of S\~{a}o Paulo,
  S\~{a}o Paulo, Brazil} 
\affiliation{Fermi National Accelerator Laboratory, Batavia, Illinois 60510}
\author{A.R.~Barker}
\affiliation{University of Colorado, Boulder, Colorado 80309}
\author{M.~Barrio}
\affiliation{The Enrico Fermi Institute, The University of Chicago, 
Chicago, Illinois 60637}
\author{L.~Bellantoni}
\affiliation{Fermi National Accelerator Laboratory, Batavia, Illinois 60510}
\author{A.~Bellavance}
\affiliation{Rice University, Houston, Texas 77005}
\author{J.~Belz}
\affiliation{Rutgers University, Piscataway, New Jersey 08854}
\author{R.~Ben-David}
\affiliation{Fermi National Accelerator Laboratory, Batavia, Illinois 60510}
\author{D.R.~Bergman}
\email[To whom correspondence should be addressed: ]{dbergman@fnal.gov}
\affiliation{Rutgers University, Piscataway, New Jersey 08854}
\author{E.~Blucher}
\affiliation{The Enrico Fermi Institute, The University of Chicago, 
Chicago, Illinois 60637} 
\author{G.J.~Bock}
\affiliation{Fermi National Accelerator Laboratory, Batavia, Illinois 60510}
\author{C.~Bown}
\affiliation{The Enrico Fermi Institute, The University of Chicago, 
Chicago, Illinois 60637} 
\author{S.~Bright}
\affiliation{The Enrico Fermi Institute, The University of Chicago, 
Chicago, Illinois 60637}
\author{E.~Cheu}
\affiliation{University of Arizona, Tucson, Arizona 85721}
\author{S.~Childress}
\affiliation{Fermi National Accelerator Laboratory, Batavia, Illinois 60510}
\author{R.~Coleman}
\affiliation{Fermi National Accelerator Laboratory, Batavia, Illinois 60510}
\author{M.D.~Corcoran}
\affiliation{Rice University, Houston, Texas 77005}
\author{G.~Corti}
\affiliation{The Department of Physics and Institute of Nuclear and 
Particle Physics, University of Virginia, 
Charlottesville, Virginia 22901} 
\author{B.~Cox}
\affiliation{The Department of Physics and Institute of Nuclear and 
Particle Physics, University of Virginia, 
Charlottesville, Virginia 22901}
\author{M.B.~Crisler}
\affiliation{Fermi National Accelerator Laboratory, Batavia, Illinois 60510}
\author{A.R.~Erwin}
\affiliation{University of Wisconsin, Madison, Wisconsin 53706}
\author{R.~Ford}
\affiliation{Fermi National Accelerator Laboratory, Batavia, Illinois 60510}
\author{A.~Glazov}
\affiliation{The Enrico Fermi Institute, The University of Chicago, 
Chicago, Illinois 60637}
\author{A.~Golossanov}
\affiliation{The Department of Physics and Institute of Nuclear and 
Particle Physics, University of Virginia, 
Charlottesville, Virginia 22901}
\author{G.~Graham}
\affiliation{The Enrico Fermi Institute, The University of Chicago, 
Chicago, Illinois 60637} 
\author{J.~Graham}
\affiliation{The Enrico Fermi Institute, The University of Chicago, 
Chicago, Illinois 60637}
\author{K.~Hagan}
\affiliation{The Department of Physics and Institute of Nuclear and 
Particle Physics, University of Virginia, 
Charlottesville, Virginia 22901}
\author{E.~Halkiadakis}
\affiliation{Rutgers University, Piscataway, New Jersey 08854}
\author{J.~Hamm}
\affiliation{University of Arizona, Tucson, Arizona 85721}
\author{K.~Hanagaki}
\affiliation{Osaka University, Toyonaka, Osaka 560-0043 Japan}  
\author{S.~Hidaka}
\affiliation{Osaka University, Toyonaka, Osaka 560-0043 Japan}
\author{Y.B.~Hsiung}
\affiliation{Fermi National Accelerator Laboratory, Batavia, Illinois 60510}
\author{V.~Jejer}
\affiliation{The Department of Physics and Institute of Nuclear and 
Particle Physics, University of Virginia, 
Charlottesville, Virginia 22901}
\author{D.A.~Jensen}
\affiliation{Fermi National Accelerator Laboratory, Batavia, Illinois 60510}
\author{R.~Kessler}
\affiliation{The Enrico Fermi Institute, The University of Chicago, 
Chicago, Illinois 60637}
\author{H.G.E.~Kobrak}
\affiliation{University of California at San Diego, La Jolla,
  California 92093} 
\author{J.~LaDue}
\affiliation{University of Colorado, Boulder, Colorado 80309}
\author{A.~Lath}
\affiliation{Rutgers University, Piscataway, New Jersey 08854}
\author{A.~Ledovskoy}
\affiliation{The Department of Physics and Institute of Nuclear and 
Particle Physics, University of Virginia, 
Charlottesville, Virginia 22901}
\author{P.L.~McBride}
\affiliation{Fermi National Accelerator Laboratory, Batavia, Illinois 60510}
\author{P.~Mikelsons}
\affiliation{University of Colorado, Boulder, Colorado 80309}
\author{E.~Monnier}
\altaffiliation[Permanent address ]{C.P.P. Marseille/C.N.R.S., France}
\affiliation{The Enrico Fermi Institute, The University of Chicago, 
Chicago, Illinois 60637}
\author{T.~Nakaya}
\affiliation{Fermi National Accelerator Laboratory, Batavia, Illinois 60510}
\author{K.S.~Nelson}
\affiliation{The Department of Physics and Institute of Nuclear and 
Particle Physics, University of Virginia, 
Charlottesville, Virginia 22901}
\author{H.~Nguyen}
\affiliation{Fermi National Accelerator Laboratory, Batavia, Illinois 60510}
\author{V.~O'Dell}
\affiliation{Fermi National Accelerator Laboratory, Batavia, Illinois 60510} 
\author{M.~Pang}
\affiliation{Fermi National Accelerator Laboratory, Batavia, Illinois 60510} 
\author{R.~Pordes}
\affiliation{Fermi National Accelerator Laboratory, Batavia, Illinois 60510}
\author{V.~Prasad}
\affiliation{The Enrico Fermi Institute, The University of Chicago, 
Chicago, Illinois 60637}
\author{X.R.~Qi}
\affiliation{Fermi National Accelerator Laboratory, Batavia, Illinois 60510} 
\author{B.~Quinn}
\affiliation{The Enrico Fermi Institute, The University of Chicago, 
Chicago, Illinois 60637} 
\author{E.J.~Ramberg}
\affiliation{Fermi National Accelerator Laboratory, Batavia, Illinois 60510} 
\author{R.E.~Ray}
\affiliation{Fermi National Accelerator Laboratory, Batavia, Illinois 60510}
\author{A.~Roodman}
\affiliation{The Enrico Fermi Institute, The University of Chicago, 
Chicago, Illinois 60637} 
\author{M.~Sadamoto}
\affiliation{Osaka University, Toyonaka, Osaka 560-0043 Japan} 
\author{S.~Schnetzer}
\affiliation{Rutgers University, Piscataway, New Jersey 08854}
\author{K.~Senyo}
\affiliation{Osaka University, Toyonaka, Osaka 560-0043 Japan} 
\author{P.~Shanahan}
\affiliation{Fermi National Accelerator Laboratory, Batavia, Illinois 60510}
\author{P.S.~Shawhan}
\affiliation{The Enrico Fermi Institute, The University of Chicago, 
Chicago, Illinois 60637}
\author{J.~Shields}
\affiliation{The Department of Physics and Institute of Nuclear and 
Particle Physics, University of Virginia, 
Charlottesville, Virginia 22901}
\author{W.~Slater}
\affiliation{University of California at Los Angeles, Los Angeles,
  California 90095} 
\author{N.~Solomey}
\affiliation{The Enrico Fermi Institute, The University of Chicago, 
Chicago, Illinois 60637}
\author{S.V.~Somalwar}
\affiliation{Rutgers University, Piscataway, New Jersey 08854} 
\author{R.L.~Stone}
\affiliation{Rutgers University, Piscataway, New Jersey 08854} 
\author{E.C.~Swallow}
\affiliation{The Enrico Fermi Institute, The University of Chicago, 
Chicago, Illinois 60637} 
\affiliation{Elmhurst College, Elmhurst, Illinois 60126}
\author{S.A.~Taegar}
\affiliation{University of Arizona, Tucson, Arizona 85721}
\author{R.J.~Tesarek}
\affiliation{Rutgers University, Piscataway, New Jersey 08854} 
\author{G.B.~Thomson}
\affiliation{Rutgers University, Piscataway, New Jersey 08854}
\author{P.A.~Toale}
\affiliation{University of Colorado, Boulder, Colorado 80309}
\author{A.~Tripathi}
\affiliation{University of California at Los Angeles, Los Angeles,
  California 90095} 
\author{R.~Tschirhart}
\affiliation{Fermi National Accelerator Laboratory, Batavia, Illinois 60510}
\author{S.E.~Turner}
\affiliation{University of California at Los Angeles, Los Angeles,
  California 90095} 
\author{Y.W.~Wah}
\affiliation{The Enrico Fermi Institute, The University of Chicago, 
Chicago, Illinois 60637}
\author{J.~Wang}
\affiliation{University of Arizona, Tucson, Arizona 85721}
\author{H.B.~White}
\affiliation{Fermi National Accelerator Laboratory, Batavia, Illinois 60510} 
\author{J.~Whitmore}
\affiliation{Fermi National Accelerator Laboratory, Batavia, Illinois 60510}
\author{B.~Winstein}
\affiliation{The Enrico Fermi Institute, The University of Chicago, 
Chicago, Illinois 60637} 
\author{R.~Winston}
\affiliation{The Enrico Fermi Institute, The University of Chicago, 
Chicago, Illinois 60637} 
\author{T.~Yamanaka}
\affiliation{Osaka University, Toyonaka, Osaka 560-0043 Japan}
\author{E.D.~Zimmerman}
\affiliation{The Enrico Fermi Institute, The University of Chicago, 
Chicago, Illinois 60637} 

\collaboration{The KTeV Collaboration}

\begin{abstract}
  We present a new measurement of the branching ratio of the decay
  \Kpeng\ (\kpeng) with respect to \Kpen\ (\kpen), and the first
  study of the photon energy spectrum in this decay.  We find
  $\BR(\kpeng, \Egs>30{\rm\ MeV}, \thegs>20^\circ)/\BR(\kpen) = 0.908
  \pm0.008{\rm (stat.)}^{+0.013}_{-0.012}{\rm (syst.)}\%$.  Our
  measurement of the spectrum is consistent with inner bremsstrahlung
  as the only source of photons in \kpeng.
\end{abstract}

\maketitle

\section{Introduction}

Precise measurements of the properties of radiative decay modes of
kaons test theories of kaon structure.  Additionally, understanding
the phenomenology of these decays enhances the ability to perform
other precision measurements and rare decay searches which rely on
robust background predictions of which radiative decays may be a
component.  There are two distinct components in most radiative
decays: direct emission (DE), where the photon is irreducibly part of
the decay interaction; and inner bremsstrahlung (IB), where the photon
is emitted from an external charged leg.  The IB component is well
understood and dominant in most radiative decays.  It is the size and
structure of the DE component which is important for understanding
kaon structure.

Fearing, Fishbach and Smith (FFS)\cite{FFS:1969,FFS:1970a,FFS:1970b}
and Doncel\cite{Doncel:1970} performed the first theoretical studies
of radiative \kpen\ decays (\Kpeng, \kpeng).  FFS present the matrix
element with the IB component and a phenomenological model of the DE
component:
\begin{equation}
  \begin{split}
    T(\kpeng) = &T_{\rm IB}+\\
                &\frac{A}{M^2}
                  (\epsilon\cdot l K\cdot k - \epsilon\cdot K l\cdot k)+\\
                &\frac{B}{M^2}(\epsilon_{\mu\nu\alpha\beta}
                  \epsilon^\mu l^\nu K^\alpha k^\beta)+\\
                &\frac{C}{M^2}
                  (\epsilon\cdot l Q\cdot k - \epsilon\cdot Q l\cdot k)+\\
                &\frac{D}{M^2}(\epsilon_{\mu\nu\alpha\beta}
                  \epsilon^\mu l^\nu Q^\alpha k^\beta)
  \label{eq:ffs}
  \end{split}
\end{equation}
where $T_{\rm IB}$ is the IB portion of the matrix element, $M$ is
the kaon mass, $\epsilon$ is photon polarization, $l$ is the
electron-neutrino current vector, $k$ is the photon momentum, $K$ is
the kaon momentum and $Q$ is the pion momentum.  They present a
calculation of the relative \kpeng\ branching ratio for photon
energies above 30 MeV in the kaon center of momentum frame (CM), using
current algebra to estimate the size of the four parameters in
Equation~\eqref{eq:ffs}.  FFS estimate a DE component roughly 1\% the
size of the IB component, but do not include it in their prediction of
the \kpeng\ branching ratio.  Doncel performs a similar calculation of
only the IB component, emphasizing in addition the need to cut on the
angle between the electron and photon in the CM in order to avoid a
background from bremsstrahlung photons generated in the detector
material\cite{Doncel:1970}.

More recently, Holstein\cite{Holstein:1990} and Bijnens \emph{et
al.}\cite{Bijnens:1993} have performed Chiral Perturbation Theory
($\chi$PT) calculations.  Holstein finds a smaller DE component than
FFS, by a factor of 5--10, but like FFS does not include it in a
numerical prediction of the branching ratio.  His prediction for the
branching ratio is the same as FFS.  Bijnens \emph{et al.} do not
explicitly separate their calculation into IB and DE pieces.

The NA31 collaboration\cite{Leber:1996} performed the most recently
published measurement of the \kpeng\ branching ratio.  With
approximately 1400 events, they measured a radiative fraction
$\BR(\kpeng, \Egs\ge 30\ \MeV , \thegs\ge20^\circ)/\BR(\kpen) =
0.934\pm0.036^{+0.055}_{-0.039}\%$.  With over 15000 \kpeng\ events,
KTeV approaches the sensitivity required to see the DE component in
this mode.

\section{The KTeV Detector}

\begin{figure} 
  \includegraphics[width=\columnwidth]{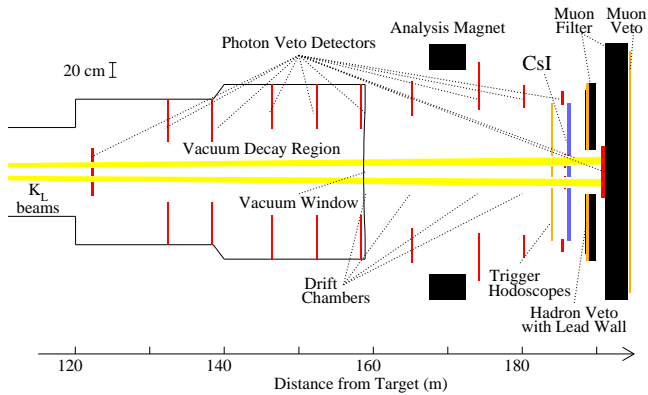}
  \caption{The KTeV Detector, as configured for collecting the data
    used in this analysis.}
  \label{fig:ktev} 
\end{figure}

This measurement was performed using the KTeV detector at Fermilab,
shown in Figure~\ref{fig:ktev}.  KTeV was designed to measure
$\Re(\epsilon'/\epsilon)$, the direct component of CP violation in neutral
kaon decays, which requires excellent detection capabilities of both
charged and neutral particles.  Hence, KTeV is an ideal apparatus for
studying decays like \kpeng\ which have both charged and neutral
particles in the final state.  The aspects of the apparatus pertinent
to this analysis are described below.

The $K_L$ beams used in this experiment were produced by an 800 GeV
beam of protons striking a beryllium oxide target.  Collimators and
sweeping magnets downstream of the target produced two, nearly
parallel neutral beams.  The decay volume for accepted decays began
110 meters from the target to allow the $K_S$ component to decay away,
and continued to the first drift chamber of the spectrometer at 160
meters.  The acceptance for decays upstream of 122 meters was
restricted by the ``Mask Anti'' (MA) anti-coincidence counter which
had holes to allow the beams to pass.  The kaon beams and decay
products traveled through vacuum from the target to the first drift
chamber.

The KTeV spectrometer measured the charged decays products.  It
consisted of four rectangular drift chambers, each with two horizontal
and two vertical planes of sense wires, and a large dipole magnet
which imparted a transverse momentum of 0.412 GeV/$c$.  The drift
chambers measured horizontal and vertical track position with a
resolution of 110 $\mu$m and momentum with a resolution of 0.4\% at a
typical momentum of 36 GeV/$c$.

Energy measurements and particle identification were performed by a
pure cesium iodide electromagnetic calorimeter (CsI).  It consisted of
3100 blocks in a square array 1.9 m on a side and 0.5 m deep.  Two 15
cm square beam holes allowed the passage of the neutral beams through
the calorimeter.  The calorimeter was calibrated using
momentum-analyzed electrons, with average energy resolution for
electrons of 0.75\% (where the momentum resolution has not been
subtracted out).  The calorimeter was also used to associate tracks
between vertical and horizontal views.

Following the CsI was a five meter long iron muon filter.  Muons
needed to have a momentum greater than 7 GeV/$c$ in order to traverse
the filter and deposit energy in the veto scintillation plane.

A series of ``photon veto'' counters surrounded the fiducial volume of
the detector to detect particles missing the CsI.  These veto counters
suppressed the background from \Kpmz\ (\kpmz) and other decays with
extra photons.

The data for this analysis were collected in a dedicated low intensity
(10\% of nominal $\epsilon'/\epsilon$ intensity) run during a 24 hour
period in 1997.  The low intensity of these runs was crucial to having
a manageable accidental background component to our measurement.  Only
minimal trigger conditions were applied: trigger hodoscope signals
indicating the presence of at least two charged tracks, in
anti-coincidence with photon veto counters.

\section{Analysis}

The analysis of this decay proceeded in two stages.  First, an
inclusive \kpen\ sample was isolated, then a subsample containing one
photon was identified.

\subsection{\kpen\ Sample}

\begin{figure}
  \includegraphics[width=\columnwidth]{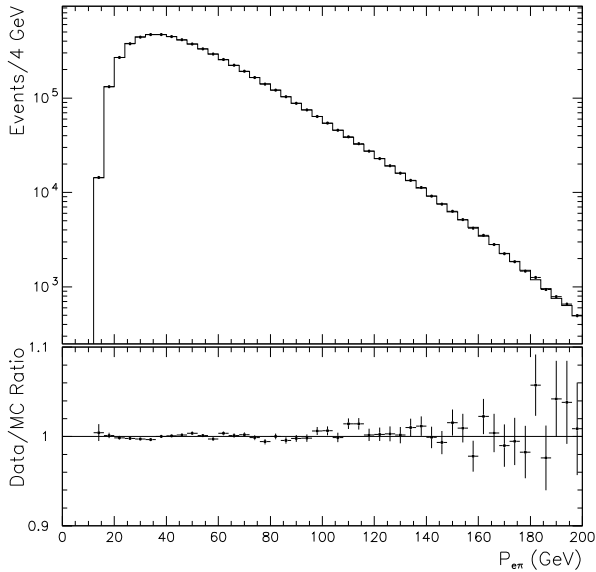}
  \caption{Data/MC comparisons of the total charged momenta
    ($P_{e\pi}$).  The $\chi^2/{\rm DOF}=95.5/72.$} 
  \label{fig:momenta_ep}
\end{figure}

The presence of an electron and a pion originating from a common
vertex define the \kpen\ normalization sample.  Particle
identification was performed by comparing the momentum measured by the
spectrometer with the energy deposited in the calorimeter: the
electron was required to have $0.95<E/P<1.1$; the pion, $E/P<0.7$.
Events with extra tracks were rejected.  The momentum of each track
was required to be above 7 GeV in order to reject muons that stop in
the muon filter before reaching the muon veto counter.  Fiducial cuts
were applied to tracks and clusters to restrict particles to well
understood parts of the detector.  Activity in the photon veto
counters was not allowed (deposited energy $<0.3$ GeV), however, there
was no cut on extra energy deposits in the CsI and photons could go
down the beam hole without being detected.  A plot of the charged
particle momentum of the normalization \kpen\ sample is shown in
Figure~\ref{fig:momenta_ep} along with a Monte Carlo simulation of the
same.

\subsubsection{The Quadratic Ambiguity}

We need to know the photon momentum in the CM for both the branching
ratio and spectrum measurements.  However, because the neutrino in
\kpen\ decays is unobserved, events cannot be reconstructed
unambiguously.  Kinematic constraints allow one to determine the
magnitude of the neutrino's CM momentum along the kaon flight
direction but not the sign (forward or back).  This leads to a
two-fold ambiguity in determining the kaon's lab momentum, and hence
the photon's CM momentum.

The acceptances for the two kaon lab momentum solutions are different,
favoring one solution over the other.  We picked the more likely
solution for any pair of momenta based on MC calculations of the
acceptance.  This procedure resulted in the correct solution 65\% of
the time.

The quadratic ambiguity is smallest in cases where the neutrino
emerges perpendicular to the kaon flight direction.  Because of the
finite resolution of the spectrometer, $\cos^2(\theta_{K\nu}^*)$ can
have non-physical, negative values.  However, events in which the
kaons scatter in the collimators can also have negative values of
$cos^2(\theta_{K\nu}^*)$.  To retain most of the unambiguous events
while rejecting large, unphysical values from events in which the kaon
scattered, we required $\cos^2(\theta_{K\nu}^*)>-0.2$.

\subsubsection{Kinematic Cuts and Backgrounds}

To increase the purity of the \kpen\ sample, we required $P_T^2 <
0.055$ GeV$^2$ and $M_{e\pi} < 0.5$ GeV, the kinematically allowed
values.  

\kpmz\ events were a background at the tenth of a percent level in the
\kpen\ sample because of the small probability (0.3\%) for a pion to
be misidentified as an electron.  \kpmz\ events were more significant
(a few percent) for the \kpeng\ subsample because the decay also has
photons present.  The mass of the unseen $\pi^0$ in \kpmz\ decays,
restricts the kinematics of the charged particles.  By eliminating
events in which the unobserved $\pi^0$ would have a physically allowed
momentum when analyzed as \kpmz\, we removed 99\% of this background,
while removing only 7\% of the \kpeng\ signal.  

Likewise, well over 99\% of \Kpm\ events with a misidentified pion
were removed by cuts on the two pion invariant mass (494--502 MeV) and
transverse momentum ($<50$ MeV$^2$).  

A few misreconstructed \kpmz\ decays remained in the sample and were
subtracted using a MC sample of \kpmz\ events.  This sample was
normalized to the data in the case where the kinematic requirement
above has been dropped and the signal had been estimated with a sample
of MC \kpen\ events.  A small \kppen\ background was also subtracted
using a MC sample normalized to the inferred data kaon flux.  These
two subtractions together change the normalization sample by less than
0.1\%.

\subsection{\kpeng\ Subsample}

The \kpeng\ subsample was selected from events in the inclusive \kpen\ 
sample.  The subsample consisted of events with exactly one photon
candidate cluster.

\begin{figure}
  \includegraphics[width=\columnwidth]{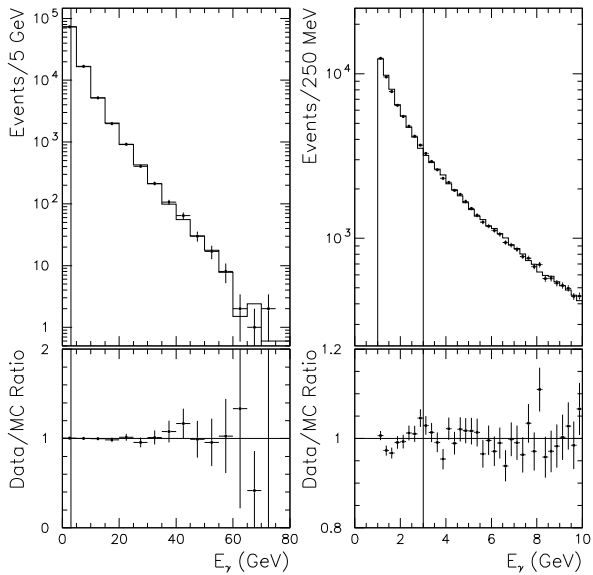}
  \caption{Data/MC comparisons of the photon energy in the laboratory
    frame at two different scales.  The photon energy was required
    to be above 3 GeV to be included in the \kpeng\ sample.  This cut
    is indicated by the vertical line in each plot.  The
    $\chi^2/{\rm DOF}=6.2/14$ for the first plot and $24.4/27$ for
    the second.}
  \label{fig:momenta_g}
\end{figure}

The photon candidate cluster was required to be more than 8 cm from
the electron cluster in the CsI and more than 40 cm from the pion
cluster.  The photon-electron distance cut allows for reliable
separation of the electron and photon clusters.  The photon-pion
distance cut removed events in which clusters due to pion shower
fluctuations, which are not well modeled in MC, might have mimicked a
photon shower.  The energy of a photon candidate cluster was required
to exceed 3 GeV, while the transverse profile of the cluster was
required to be consistent with the hypothesis of a single,
electromagnetic shower.  A plot of the photon energy in the laboratory
frame is shown in Figure~\ref{fig:momenta_g}.  The photon cluster was
also required to be more than 2 cm away from the electron's position
as projected from the upstream track segment to the CsI.  This cut
removed photons from physical bremsstrahlung as the electron passed
through the detector.

The \kpeng\ events were required to have exactly one photon candidate
cluster.  In addition, the calorimeter was required to be free of
other electromagnetic clusters with energy greater than 1 GeV and
which were separated from the electron cluster by more than 4 cm and
from the pion cluster by more than 30 cm.  The MC does not simulate
the multiplicity of these clusters perfectly.  A small correction,
$C_V$, of 0.25\% was applied to account for this.  The correction was
determined by finding the ratio of the number of events with two or
more photon candidate clusters to the number of events with one or
more photon candidate clusters.

For the branching ratio measurement, additional kinematic cuts on the
energy, $E_\gamma^*> 30$ MeV, and the angle between the electron and
photon in the CM, $\theta_{e\gamma}^*> 20^\circ$, were used in order
to obtain a result with a kinematic range commensurate with the
previous experimental result and to emphasize the interesting
kinematic region for DE studies.

In our study of the photon CM spectrum (see Section
~\ref{sec:spectrum}), the cut on the distance between the photon
candidate and the electron projection was lowered to 1 cm, to allow us
to use looser kinematic cuts $E^*> 25$ MeV and $\theta_{e\gamma}^*>
5^\circ$.

The background subtractions mentioned above for the inclusive \kpen\ 
sample were more significant in the \kpeng\ subsample, comprising
0.7\% of the total.

\section{Monte Carlo Simulation}

The detector acceptance used in calculating the branching ratio and in
determining the acceptance-corrected photon spectrum was calculated by
MC simulation.  As mentioned above, the photon momentum in \kpeng\ 
cannot be constrained due to the unseen neutrino.  This ambiguity
makes precise MC simulation essential to the success of this analysis.
To accurately simulate accidental activity in the detector, a special
trigger was used to record events at random times with a frequency
proportional to the overall beam intensity.  These random events were
overlaid on MC generated events for comparisons with the data and for
acceptance calculations.  We were aided by the low beam intensity
conditions under which the data were recorded.

FFS\cite{FFS:1970b} present a full listing of all terms for the
squared matrix element, including both IB and DE components.  We used
their listing explicitly in our Monte Carlo (MC) simulations.  For the
branching ratio measurement, all the DE coefficients were set to zero.
The MC \kpeng\ sample was combined with a MC sample of non-radiative
\kpen\ decays at a level corresponding to NA31's branching ratio
measurement\cite{Leber:1996}.  The cutoff for generating a physical
photon to trace through the detector was 1 keV in the CM.  Photons
below this energy as well as loop effects were taken into account in
the \kpen\ generation.  The \kpen\ form factor, used for both
radiative and non-radiative events was
$\lambda_+=0.0274(14)$\cite{Tesarek:1999}, based on a preliminary KTeV
measurement.  Our result depends only weakly on the form factor, and
the form factor used is in agreement with the 2000 PDG
value\cite{PDG:2000}.

Two other features of the MC simulation are worth mentioning.  First,
electromagnetic and hadronic showers in the CsI were handled by
separate libraries of showers generated with GEANT.  Second, the
tracking of particles through the detector and beamline accounted for
a number of effects including multiple scattering, bremsstrahlung and
synchrotron radiation, delta particle emission, photon conversion and
kaon regeneration.

\section{Branching Ratio Calculation}

The measured relative branching ratio is simply defined by the ratio
of the events we measure in each sample corrected by the acceptance in
each case.  It is important to realize that our dependence on MC is
mitigated by the ratio of acceptances appearing in the calculation.
The sample sizes and acceptances are listed in
Table~\ref{tab:samples}.  Thus we find,
\begin{multline}
        \frac{\Gamma(\kpeng,
                E_\gamma^*>30{\rm\ MeV},
                \theta_{e\gamma}^*>20^\circ)}
              {\Gamma(\kpen)} \\
             = \frac{\Num(\kpeng)}{\Num(\kpen)}
                 \frac{\Acc(\kpen)}{\Acc(\kpeng)}
                 C_V \\
             = 0.908\pm0.008({\rm stat})^{+0.013}_{-0.012}({\rm syst})\%
\end{multline}
As mentioned above, the correction, $C_V=1.0025\pm0.0009$, is
necessary to take into account the different rates at which \kpeng\ 
candidate events are vetoed by the cluster multiplicity cut in data
and MC.  Vetoed events in the MC are not included in the MC
calculation of the acceptance and thus require an explicit
correction.

\begin{table}
\begin{displaymath}
\begin{array}{lr}\hline\hline
\multicolumn{2}{c}{{\rm Normalization,\ } \kpen} \\ \hline
{\rm Raw Events}                  &5760888\\
\qquad\kpmz{\rm \ background}     &    267\\
\qquad\kppen{\rm \ background}    &    197\\
{\rm Background subtracted events}&5760424\\
\kpen {\rm \ Acceptance}           &0.1067\\\hline
\multicolumn{2}{c}{{\rm Signal,\ } \kpeng(E_\gamma^*>30{\rm\ MeV},
                          \theta_{e\gamma}^*>20^\circ)} \\ \hline 
{\rm Raw Events}                  &  15575\\
\qquad\kpmz{\rm \ background}     &     50\\
\qquad\kppen{\rm \ background}    &     62\\
{\rm Background subtracted events}&  15463\\
\kpeng {\rm  Acceptance}          &0.03161(14)\\\hline\hline

\end{array}
\end{displaymath}
\caption{Sample sizes used in the branching ratio calculation.}
\label{tab:samples}
\end{table}

\begin{figure}
  \includegraphics[width=\columnwidth]{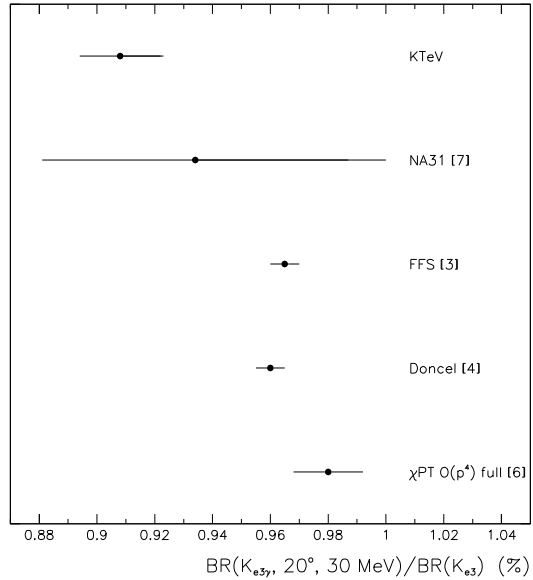}
  \caption{A comparison of our branching ratio measurement with other
    recent measurements and predictions.}
  \label{fig:brcomp}
\end{figure}

Our result is compared with the other published measurements and
predictions in Figure~\ref{fig:brcomp}.  The uncertainties for the
theoretical predictions are based solely on the accuracy of the stated
results.  FFS give a result only for $\theta_{e\gamma}^*> 0$.  We have
used our MC (based explicitly on the FFS~\cite{FFS:1970b} matrix
element) to calculate the fraction of the events with
$\theta_{e\gamma}^*> 20^\circ$ and give the result for this case.

\subsection{Systematic Uncertainties}

Estimates of the sizes of various significant systematic uncertainties
are listed in Table~\ref{tab:systematics}.  The largest systematic
uncertainty comes from the variation of the calculated branching ratio
with the $E_\gamma$ cut, as the cut was varied by $\pm1$ GeV about 3
GeV.  Photons below 2 GeV are not well reconstructed as we have seen
by analyzing large samples of \kpmz\ decays.  The relative size of
this uncertainty reflects the steeply falling photon spectrum at low
energies combined with the difficulty of modeling the calorimeter at
the lowest energies.

The error associated with picking the right momentum solution was
based on the degree of consistency between the subset of the data
where momentum solution is unambiguous within resolution, and the data
set as a whole.  The error on the acceptance was based on the
statistics of MC generated.  To this was added the variation in
acceptance due to the unknown DE parameters studied below.  The upper
limit on the background due to clusters originating from pion shower
fluctuations was determined by studying \kpmn\ decays in data.

The contribution of accidental activity in the CsI to the \kpeng\ 
subsample was small, 0.5\%, as determined by generating MC with and
without accidental overlays and comparing the number of events with
exactly one photon candidate.  The number of events with more than one
photon candidate agreed between data and MC with accidental overlays
at the 20\% level, whereas there were differences of orders of
magnitude without accidental overlays.

Other errors were estimated from the differences between data and MC
shapes of various distributions combined with different expected
shapes between normalization and signal samples, and the uncertainty
in the measured \kpen\ form factor.

\begin{table}
\begin{displaymath}
\begin{array}{lrr}\hline\hline
                      & {\rm Down (\%)} & {\rm Up (\%)}\\\hline
  E_\gamma {\rm \ cut}                  & -0.005 & +0.008\\
  K\ {\rm momentum solution}            & -0.007 & +0.007\\
  {\rm MC Statistics}                   & -0.006 & +0.006\\
  {\rm Pion shower BG}                  & -0.003 & +0.000\\
  \kpen\ {\rm form factor}              & -0.003 & +0.003\\
  P_e^* {\rm \ shape}                   & -0.003 & +0.003\\
  P_T^2 {\rm \ shape}                   & -0.002 & +0.002\\
  {\rm Veto Correction}                 & -0.001 & +0.001\\
  {\rm Photon Conversions}              & -0.001 & +0.001\\
  \cos^2(\theta_{K\nu}^*) {\rm \ shape} & -0.001 & +0.001\\
  {\rm Accidental Overlay}              & -0.001 & +0.001\\
  \kpmz\ {\rm BG subtraction}           & -0.001 & +0.001\\
  \kppen\ {\rm BG subtraction}          & -0.001 & +0.001\\\hline
  {\rm Total Systematic Uncertainty}    & -0.012 & +0.013\\
  {\rm Statistical Uncertainty}         & -0.008 & +0.008\\\hline
  {\rm Total Uncertainty}               & -0.014 & +0.015\\\hline\hline
\end{array}
\end{displaymath}
\caption{A list of significant uncertainties in the branching ratio
  measurement.} 
\label{tab:systematics}
\end{table}

\section{Spectrum Measurement}
\label{sec:spectrum}

With the statistics available in this experiment, it becomes feasible
to compare the FFS model for direct emission (Eq.  \eqref{eq:ffs})
with our observed photon CM momentum spectrum.  We do not have the
sensitivity to consider all four parameters, so we use the soft-kaon
approximation\cite{Holstein:1999} to set $A=B=0$ and restricted our
study to $C$ and $D$ whose terms remain non-zero in this
approximation.  The soft-kaon approximation assumes that terms in the
decay matrix element proportional to the kaon rest mass are
negligible.

To determine the values of $C$ and $D$ favored by our data, we used MC
to generate spectra at a set of points in $CD$ space.  We compared
these spectra, bin-by-bin, to the acceptance-corrected spectrum seen
in the data to calculate a $\chi^2$ value at each point.  The only
free parameter in this comparison was the normalization of the two
spectra.  To increase the statistics for these fits the
photon-electron angle cut was relaxed to $\theta_{e\gamma}^*>
5^\circ$.  This cut is still far from allowing physical bremsstrahlung
events into our sample.  The fit was done for photon energies between
25 and 200 MeV in seven bins.  Below 25 MeV we were limited by the
sensitivity of our calorimeter to soft photons.  Above 200 MeV, the
error due to making the wrong kaon momentum choice became dominant.

\begin{figure}
  \includegraphics[width=\columnwidth]{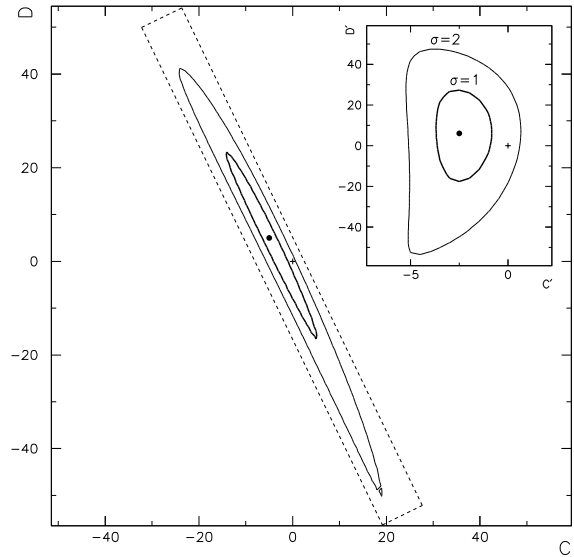}
  \caption{The $\sigma=1$ \mbox{($\chi^2-\chi_{\rm min}^2=1$)} and
    $\sigma=2$ \mbox{($\chi^2-\chi_{\rm min}^2=4$)} contours of the
    fit/interpolated $\chi^2$ surface as a function of $C$ and $D$.
    The dot shows the position of the minimum, while the cross shows
    the IB-only point.  The region within the oblique rectangle is
    rotated and redisplayed in the inset.  The new axes $C'$ and $D'$
    are obtained from $C$ and $D$ after a rotation of
    $\theta=25.8^\circ$.}
  \label{fig:gcdppqtfcd}
\end{figure}

The constant $\chi^2$ contours corresponding to $\sigma=1$
\mbox{($\chi^2-\chi_{\rm min}^2=1$)} and $\sigma=2$
\mbox{($\chi^2-\chi_{\rm min}^2=4$)} as a function of $C$ and $D$
are shown in Figure~\ref{fig:gcdppqtfcd}.  A polynomial interpolation
was done to estimate the values $\chi^2$ between sampled points.  We
find $C = -5\pm{10}$ and $D = 5^{+19}_{-21}$, with a strong
correlation between the two parameters.

We are most sensitive to the linear combination of $C$ and $D$ which
approximates a linear perturbation of the IB-only spectrum.  The
precision of our result is more apparent when one chooses axes
aligned with this combination.  Choosing axes $C'$ and $D'$ rotated
by $\theta=25.8^\circ$, we find $C' = -2.5^{+1.5}_{-1.0}$ and $D' =
6^{+22}_{-24}$.  This is illustrated in the inset of
Figure~\ref{fig:gcdppqtfcd}.

\begin{figure}
  \includegraphics[width=\columnwidth]{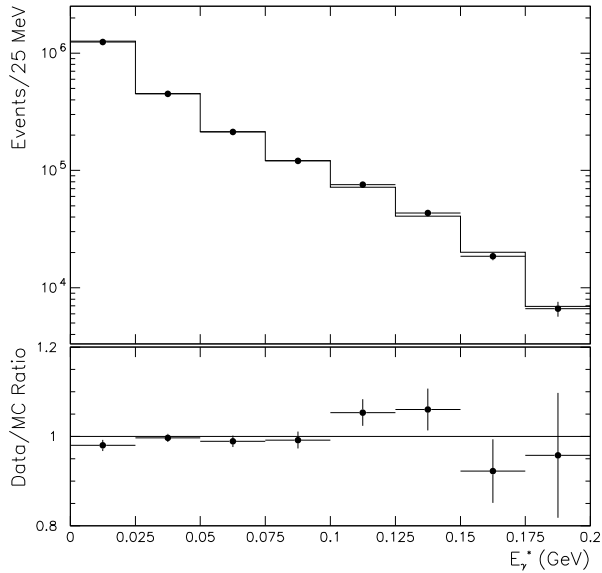}
  \caption{Data/MC comparison of the acceptance-corrected $E_\gamma^*$
    spectrum at the best $CD$ point.  The $\chi^2/{\rm DOF}=4.3/4$,
    where the 0--25~MeV bin has not been included in the comparison.}
  \label{fig:gcdppqtfch2}
\end{figure}

The observed acceptance-corrected $E_\gamma^*$ spectrum is shown in
comparison with the generated spectrum at the best-fit lattice point
in Figure~\ref{fig:gcdppqtfch2}.  The $\chi^2$ value for this point is
4.3 with four degrees of freedom.  At the IB-only point (0,0), the
$\chi^2$ is 7.4.

Knowledge of the acceptance correction is the dominant systematic
error contributing to our measurement of $C'$ and $D'$.  Physical
backgrounds are very low and insignificant in the spectrum
measurement.  We estimated the systematic uncertainty associated with
$C'$ (the large statistical uncertainty in $D'$ makes the systematic
uncertainty insignificant) by multiplying the measured acceptance
profile by a linear factor and renormalizing to keep the overall
acceptance constant.  The allowed size of the linear factor was
determined by statistical consistency with the measured acceptance.
By varying the acceptance in this manner we find a systematic
uncertainty in $C'$ of 1.5 and in $D'$ of 1.0.

\section{Conclusion}

We have measured the relative branching ratio for radiative \kpen\ 
decays and fit the photon spectrum to place limits on the size of DE
components in the \kpeng\ matrix element.  Our measurement,
$\BR(\kpeng, \Egs>30{\rm\ MeV}, \thegs>20^\circ)/\BR(\kpen) =
0.908\pm0.008({\rm stat})^{+0.013}_{-0.012}({\rm syst})\%$, is nearly
five times more precise than the previous measurement, and in
agreement with it.  It is significantly lower than all published
theoretical predictions.

This is the first attempt to measure DE terms by studying the photon
spectrum.  The spectrum measurement is consistent with IB as the only
source of photons in \kpeng\ decays.  In the soft-kaon approximation,
we find $C = -5\pm{10}$ and $D = 5^{+19}_{-21}$, or, in a frame
rotated by $25.8^\circ$, $C' = -2.5^{+1.5}_{-1.0}({\rm stat})
\pm1.5({\rm syst})$ and $D' = 6^{+22}_{-24}\pm1.0({\rm syst})$.

\begin{acknowledgments}
  We gratefully acknowledge the support and effort of the Fermilab
  staff and the technical staffs of the participating institutions for
  their vital contributions.  This work was supported in part by the
  U.S.  Department of Energy, The National Science Foundation and The
  Ministry of Education and Science of Japan.  In addition, A.R.B.,
  E.B. and S.V.S.  acknowledge support from the NYI program of the
  NSF; A.R.B. and E.B. from the Alfred P. Sloan Foundation; E.B. from
  the OJI program of the DOE; K.H., T.N. and M.S. from the Japan
  Society for the Promotion of Science; and R.F.B from the
  Funda\c{c}\~{a}o de Amparo \`{a} Pesquisa do Estado de S\~{a}o
  Paulo.  P.S.S. acknowledges receipt of a Grainger Fellowship.
\end{acknowledgments}

\end{document}